\documentclass[%
 reprint,
 amsmath,amssymb,
 aps,
]{revtex4-2}
\usepackage{graphicx}% Include figure files
\usepackage{dcolumn}% Align table columns on decimal point
\usepackage{bm}% bold math
\usepackage{url}
\usepackage{hyperref}
\usepackage[dvipsnames]{xcolor}
\definecolor{blue(pigment)}{rgb}{0.2, 0.2, 0.6}
\hypersetup{
    allcolors=blue(pigment),
    colorlinks=true,
    linkcolor=blue(pigment),
    filecolor=blue(pigment),      
    urlcolor=blue(pigment),
}

\newcommand{\Rnum}[1]{\lowercase\expandafter{\romannumeral #1\relax}}
\newcommand{\RNum}[1]{\uppercase\expandafter{\romannumeral #1\relax}}

\begin{document}

\preprint{APS/123-QED}

\title{Sensing Quantum Nature of Primordial Gravitational Waves Using Electromagnetic Probes}% Force line breaks with \\


\author{F. Shojaei Arani}
\email{fateme.shojaei@irap.omp.eu}
\affiliation{%
Department of Physics, University of Isfahan, Hezar Jerib Str., Isfahan 81746-73441, Iran;\\
}%
\affiliation{%
Université de Toulouse, UPS-OMP, IRAP, F-31400 Toulouse, France
}%
\affiliation{%
CNRS, IRAP, 14, avenue Edouard Belin, F-31400 Toulouse, France
}%

\author{M. Bagheri Harouni}
\email{m.bagheri@sci.ui.ac.ir}
\affiliation{
Department of Physics, University of Isfahan, Hezar Jerib Str., Isfahan 81746-73441, Iran;\\
}%
\affiliation{
Quantum Optics Group, Department of Physics, University of Isfahan, Hezar Jerib Str., Isfahan 81746-73441, Iran;\\
}%

\author{Brahim Lamine, and Alain Blanchard}
\email{brahim.lamine@irap.omp.eu}
\affiliation{%
Université de Toulouse, UPS-OMP, IRAP, F-31400 Toulouse, France
}%
\affiliation{%
CNRS, IRAP, 14, avenue Edouard Belin, F-31400 Toulouse, France
}%


\date{\today}% It is always \today, today,

\begin{abstract}
Based on optical medium analogy, we establish a formalism to describe the interaction between an electromagnetic (EM) system with gravitational waves (GWs) background. After a full discussion on the classical treatment of the EM-GW interaction and finding the EM field mode-functions in the presence of the magneto-dielectric media caused by GWs, the governing quantum interaction Hamiltonian is obtained. Investigation of the optical quadrature variance as well as the visibility of a laser field interacting with the multi-mode squeezed primordial gravitational waves imply that the inflationary primordial gravitational waves (PGWs) act as a decoherence mechanism that destroy EM coherency after a characteristic time scale, $\tau_{c}$, which depends on the inflationary parameters $(\beta,\beta_s,r)$, or equivalently, the fractional energy density of PGWs, $\Omega_{gw,0}$. The decoherency mechanism overcomes the coherent effects, such as revivals of optical squeezing, thus leaving their confirmation out of reach. Influenced by the continuum of the squeezed PGWs, the laser field suffers a line-width broadening by $\gamma= \tau_{\text{c}}^{-1}$. The most peculiar property of the EM spectrum is the apparition of side bands at $\omega\sim \omega_0\pm 1.39 \tau_c^{-1}$Hz, stemming from the squeezed nature of PGWs. The laser phase noise induced by the squeezed PGWs grows with time squarely, $\Delta\phi=(t/\tau_c)^2$, that can most possibly be sensed within a finite flight time.
\end{abstract}

%Use showkeys class option if keyword
\maketitle

%\tableofcontents

\section{\label{sec:1}Introduction}

Our findings of the early Universe cosmology are limited predominantly by the photons and neutrinos of the last scattering surface and some signatures of the inflationary scenario on the cosmic microwave background (CMB), when the Universe was opaque. However, observations of gravitational waves (GWs) ~\cite{abbott2016observation} have opened a new window to detect the exotic gravitational phenomena, using GWs as a new messenger capable to manifest the entire Universe, even the remnants of the big bang \cite{seoane2013gravitational}.

According to the inflationary theory, primordial quantum fluctuations constitute the origin of the structure of our Universe as well as the primordial tensor fluctuations. The existence of the primordial gravitational waves background (PGWs) is one of the key predictions of the inflationary scenario of the early Universe and their detection will provide us a hitherto inaccessible view of the Universe. Spanning a full range of frequency from $10^{-19}$Hz to $10^{10}$Hz, PGWs are going to be probed at different frequency windows by means of various GWs detectors such as LIGO \cite{zhao2006relic}, the planned laser interferometer space-based antenna (LISA) \cite{chongchitnan2006prospects} and by studies on the B-mode polarization of the cosmic background radiation \cite{suzuki2018litebird}. Moreover, the third generation GW observatories such as the Einstein telescope (ET) and the Cosmic Explorer (CE) with sensitivities higher than the advanced detectors in the high frequency bands ($\sim 100$Hz), has opened a new avenue to detect the stochastic background of GWs predicted by the inflationary Universe models \cite{punturo2010einstein,hall2021gravitational}.

On top of that, it is believed that PGWs generated by the strong gravitational pumping engine of the very early Universe had evolved into the so-called squeezed states, and after that, they decoupled from the rest of matter and radiation, and freely propagated throughout the Universe \cite{caprini2018cosmological}. The today's spectrum of the PGWs crucially depends on their initial amplitude at the inflationary era, as well as the whole expansion history of the Universe. Hence, not only detecting their spectrum is significant, but also revealing their quantum mechanical nature could affirm the quantum behaviour of gravity at scales much larger than the Plank's length.

Various theoretical schemes based on quantum optical techniques have been proposed to search for the non-classical nature of gravity \cite{bassi2017gravitational,marletto2017gravitationally,belenchia2019information}. In particular, several schema are considered to reveal the squeezed essence of the PGWs, using either ground- and space-based gravitational wave interferometers \cite{allen1999squeezing,bose2002observational}, or by observing their imprint on the CMB temperature fluctuations and polarizations \cite{grishchuk2010discovering,matsumura2020squeezing}, or by performing a Hanbury Brown-Twiss interferometery on the PGWs \cite{jones2019gravitational,giovannini2019quantum, kanno2019detecsting}. On the experimental side, however, the search for demonstrations of the quantum nature of PGWs is still challenging.

In the present contribution, we develop the optical medium analogy (OMA) framework to express the EM-GW interaction using the familiar quantum optical tools. This approach has been vastly used to better grasp the physical aspects of gravitational phenomena at the classical level \cite{plebanski1960electromagnetic,mashhoon1973scattering, mashhoon1979detection, mashhoon1982contribution,bini2014refraction}. In the OMA formalism, the GW background manifests itself as a non-dispersive non-absorptive magneto-dielectric media, possessing an anisotropic index of refraction. At the classical level, besides recovering the previously known basic formula for the response function of a large GW detector such as the LISA interferometer, we investigate the (\Rnum{1}) walk-off angle of light in the birefringence medium caused by GWs and (\Rnum{2}) the Stokes parameters of the EM signal.

Going forward to the quantum level, we show that the interaction Hamiltonian resembles the opto-mechanical (OM) coupling interaction, as suggested by \cite{pang2018quantum}.
The OM analogy inspires the idea of employing the equivalent full-grown field of quantum OM research, to quest for the experimental demonstration of the quantum nature of PGWs, pretty similar to what was already done to expose the quantum behavior of the mesoscopic mechanical oscillators. In this regard, it has been shown that revivals of optical squeezing in an OM system is a pure quantum effect which reveals the quantum nature of the mechanical oscillator \cite{ma1811recurrence}. The same argument has been proposed to reveal the quantum nature of a single-mode squeezed GW \cite{guerreiro2020quantum}. However, the presence of the continuum of PGWs act a heat bath that suppresses the coherent dynamics, such as the revivals of optical squeezing. The decoherency of photons induced by the Gaussian stochastic background of gravitons in thermal state has been studied by Anastopoulos et al \cite{lagouvardos2021gravitational}, where it is shown that the phase uncertainty of light grows with time linearly, $\langle\Delta\phi\rangle^2 \propto t$, and the needed time to observe the effect is typically large. We'll show that the induced phase noise by the highly squeezed PGWs grows fast with time, $\langle\Delta\phi\rangle^2=(t/\tau_c)^4$, which leads to phase uncertainty $\langle\Delta\phi\rangle \sim 10^{-8}$ after a relatively small flight time of $t \sim 1 $ sec, which is likely to be measured by laser interferometer GW detectors. However, the appearance of side bands in the spectrum of light, which is another characteristic phenomenon caused by the squeezed background of GWs, seem unlikely to be probed by today's techniques.

We perform our description in the traceless transverse (TT) gauge and keep only linear terms in $h\equiv\mathcal{O}(|h_{ij}|)$, where $h_{ij}$ is the perturbation to the flat space-time metric. In the OMA formalism, GWs enter into play by changing the mode solutions of the optical wave equation through the varying refractive index of the medium. This is similar to the way that the opto-mechanical interaction comes into play through the varying boundary conditions of the EM field inside a cavity with moving end mirror \cite{pang2018quantum}. This similarity becomes more intuitive when one uses the proper detector frame to describe GWs. In that frame, the geodesic deviation equation implies that the gravitational wave acts as a tidal force leading to  proper distance between two points to change with time, which can be interpreted as ``changing boundary condition" for the EM field.

The paper is organized as follows. Section \ref{sec:2} provides a brief review on the basics of PGWs and sets the required tools of the OMA. In section \ref{sec:3}, the Maxwell's wave equations in the presence of an arbitrary background of GWs are solved. The refractive index of the medium, the frequency shift, the walk-off and the Stokes parameters describing the polarization state of the EM field are evaluated. The model is promoted to a fully quantum mechanical one in section \ref{sec:4}, where both EM and GWs fields are treated quantum mechanically. Section \ref{sec:5} is devoted to results, discussion and prospects. We conclude the paper in section \ref{sec:6}.

\section{\label{sec:2}Preliminaries}

\subsection{\label{subsec:2.A}Primordial gravitational waves}

The weakly perturbed flat space-time metric is described by $g_{\mu\nu}=\eta_{\mu\nu}+h_{\mu\nu}$ where $\eta_{\mu\nu}$ is the Minkowsky metric with signature $(- + + +)$ and $h_{\mu\nu}$ is the perturbation metric, $|h_{\mu\nu}|\ll 1$. In the linear regime one raises and lowers the indices by $\eta_{\mu\nu}$. The linearized perturbed Einstein equations in the weak-field limit take on the form of the familiar d'Alembert's equation, which in the transverse-traceless (TT) gauge have the plane-wave solution
\begin{eqnarray}\label{eq:1}
h_{ij}(\mathbf{r},t) &=& \sum_{\gamma=+,\times} \int\frac{d^3\mathbf{K}}{(2\pi)^{3/2}} e^{\gamma} _{ij}[\hat{\mathbf{K}}] \Big( h_{\gamma}(\mathbf{K})  e^{-i(\Omega t - \mathbf{K} \cdot \mathbf{r})} \\
&+& h_{\gamma}^{\ast}(\mathbf{K}) e^{i(\Omega t - \mathbf{K} \cdot \mathbf{r})} \Big)\nonumber,
\end{eqnarray}
where $i,j$ are spatial indices and $h_{\gamma}(\mathbf{K})$ is the amplitude (strain field) of the wave with polarization tensor $e^{\gamma}_{ij}[\hat{\mathbf{K}}]$. $(\Omega/c,\mathbf{K})$ represents the 4-momentum of GWs with $|\mathbf{K}|=\Omega/c$ and $\hat{\mathbf{K}}= \mathbf{K}/|\mathbf{K}|$.
For the EM field propagating through space time, the delayed time is defined as $t_{d}(\Psi)\equiv t - \hat{\mathbf{K}}\cdot\mathbf{r}/c =t(1-\hat{\mathbf{K}}\cdot\hat{r}) = t (1-\cos\Psi)$ where $\Psi$ is the angle between the EM field and GWs propagation directions and $|\mathbf{r}|=ct$. In small GWs' detectors where the spatial length of the detector, $L$, is small compared to the GWs' wavelength, $\lambda_{gw}$, time delayed effects can be neglected by setting $e^{i\mathbf{K}\cdot\mathbf{r}}\approx 1$ in (\ref{eq:1}) and $t_{d}(\Psi)\rightarrow t$. We will use the fact that the major contribution of the PGWs background comes from the ultra-low frequency part of its spectrum, for which the time delayed effects can be disregarded. 

The TT gauge leaves only two independent polarization states $+$ and $\times$ with the property $\hat{e}^{\gamma}_{ii}= \hat{e}_{ij}^{\gamma}K^{j}=0$ and $\hat{e}_{ij}^{\gamma} \hat{e}_{ij}^{\gamma} =2$ where summation convention over repeated indices is implied. The polarization tensors can be expressed in terms of two unit vectors $(\hat{\mathbf{n}},\hat{\mathbf{m}})$ orthogonal to the propagation direction $\hat{\mathbf{K}}$ and to each other. In terms of the Euler's angles $(\Theta,\Phi)$ we write
\begin{eqnarray}\label{eq:2}
\scalebox{0.8}{$\hat{K}=$} \begin{pmatrix} 
    \scalebox{0.8}{$\sin\Theta\cos\Phi $}\\
    \scalebox{0.8}{$\sin\Theta\sin\Phi $}\\ 
    \scalebox{0.8}{$\cos\Theta $}\\
	\end{pmatrix},
	\scalebox{0.8}{$\hat{\mathbf{n}}=$}\begin{pmatrix} 
    \scalebox{0.8}{$\cos\Theta\cos\Phi $} \\
    \scalebox{0.8}{$\cos\Theta\sin\Phi $}\\ 
    \scalebox{0.8}{$-\sin\Theta $}\\
	\end{pmatrix},
	\scalebox{0.8}{$\hat{\mathbf{m}}=$}\begin{pmatrix} 
    \scalebox{0.8}{$-\sin\Phi $}\\
    \scalebox{0.8}{$\cos\Phi $}\\ 
    \scalebox{0.8}{$0 $}\\
	\end{pmatrix},
\end{eqnarray}
and the polarization tensors are written as
\begin{eqnarray}\label{eq:3}
\begin{cases}
    \hat{e}_{ij}^{+}[\hat{K}] =\hat{n}_{i}\hat{n}_{j}-\hat{m}_{i}\hat{m}_{j},\\
    \hat{e}_{ij}^{\times}[\hat{K}] =\hat{m}_{i}\hat{n}_{j}+\hat{n}_{i}\hat{m}_{j}.
  \end{cases}
\end{eqnarray}
For instance, a pure $+$ polarized monochromatic GW propagating in the z-direction produces perturbations to the flat metric as
\begin{eqnarray}\label{eq:4}
ds^2 = -c^2dt^2 + \left( 1+h_{+} \right)dx^2 + \left( 1-h_{+} \right)dy^2 + dz^2.
\end{eqnarray}

\begin{figure}[htb]
 \centering
   \includegraphics[%height=50mm,
   width=0.9
   \columnwidth]{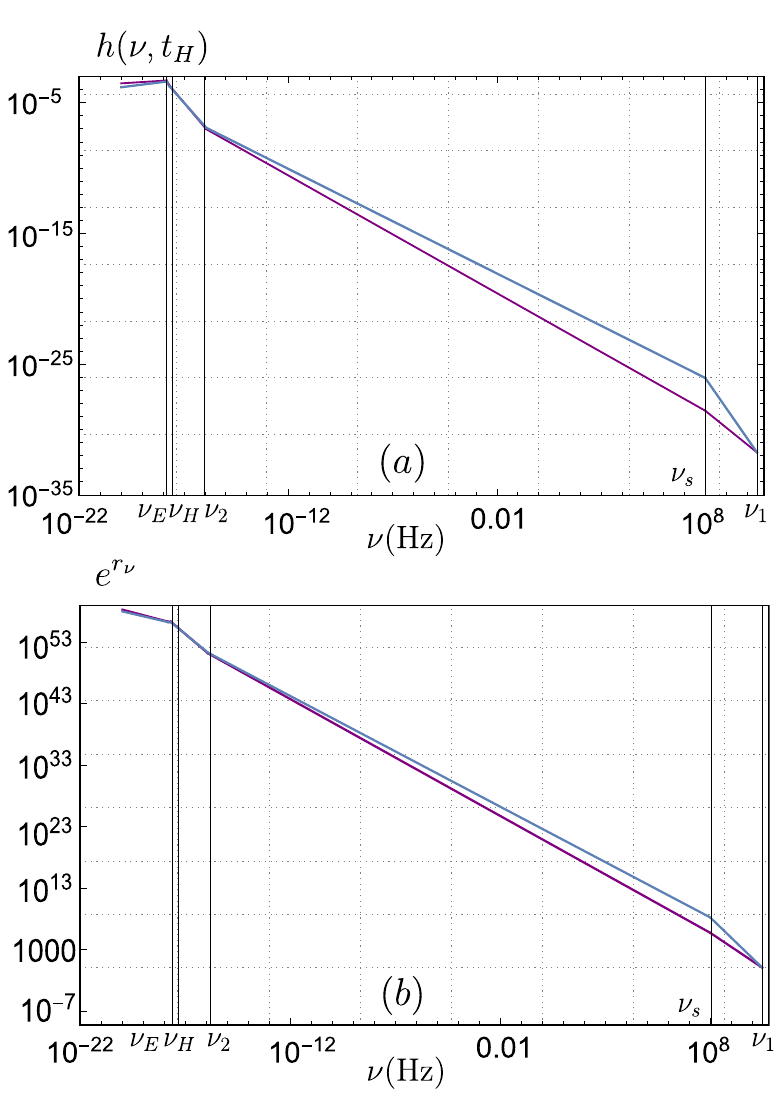}
 \caption{ (a) Today's spectrum of the PGWs and (b) the corresponding squeezing factor $e^{r_\nu}$ based on equation (\ref{eq:5}). The purple and blue curves correspond to  $(\beta,\beta_s,\mathcal{A})=(-1.9,0.38,4.9\times10^{-4})$ and $(-1.8,1.49,4.2\times10^{-4})$, respectively.}
 \label{FIG1}
\end{figure}

PGWs are small perturbations to the metric tensor of a homogeneous isotropic expanding Universe \cite{mukhanov1992theory}. Tensor perturbations of the Robertson-Walker space-time metric at the inflationary era have been subjected to the \textit{super adiabatic amplification} which enforced their corresponding vacuum state to evolve into multi-mode highly squeezed states \cite{grishchuk2010discovering}. The today's spectrum of the generated stochastic background of PGWs, $h_{\gamma}(\mathbf{K},t_{H})$, strongly depends on the cosmological model describing the expansion of the Universe as well as the initial conditions at the inflationary stage \cite{zhang2005relic,miao2007analytic,tong2009relic}. The most general models for the sequence of successive epochs of power-law expansion and the resulting tensor fluctuation spectrum are well studied in the literature \cite{zhang2005relic}, where the Universe starts from an inflationary era, followed by the reheating, radiation dominated, matter dominated and dark energy dominated stages, sequentially. Usually, the main effort is to put constraint on the model parameters ($\beta,\beta_s,r$) which describe the power-law expansion of the Universe at the inflationary and reheating epochs by $(\beta,\beta_s)$ respectively, and $r$ stands for the tensor to scalar ratio. The parameter $\beta$ characterises the spectral index of the PGWs according to $\text{n}=2\beta+5$ \cite{grishchuk2001relic}. Another relevant quantity is the squeezing amplitude of PGWs, $r_{K}$, which is defined by $e^{r_K}=a_{\ast\ast}/a_{\ast}$, where $a_{\ast}(a_{\ast\ast})$ is the value of the scale factor at $\eta_{\ast}(\eta_{\ast\ast})$, when the long-wavelength regime for a given mode $K$ begins (ends). Since we are to investigate the squeezed nature of PGWs, we mostly use the today squeezing spectrum of the whole frequency range of PGWs, specified by
\begin{eqnarray}\label{eq:5}
\hspace{-0.5cm}
e^{r_{\nu}}= \begin{cases}
\left( \frac{\nu}{\nu_{E}} \right)^{1+\beta} \left( \frac{\nu_2}{\nu_{H}} \right) \left( \frac{\nu_H}{\nu_s} \right)^{\beta} \left( \frac{\nu_s}{\nu_1} \right)^{\beta-\beta_s}, & \nu_{E}\geq \nu\\
\left( \frac{\nu}{\nu_{H}} \right)^{2+\beta} \left( \frac{\nu_2}{\nu_{H}} \right)^{1-\beta} \left( \frac{\nu_2}{\nu_s} \right)^{\beta} \left( \frac{\nu_s}{\nu_1} \right)^{\beta-\beta_s} , & \nu_{H}\geq \nu\geq \nu_{E}\\
\left( \frac{\nu_2}{\nu} \right)^{1-\beta} \left( \frac{\nu_2}{\nu_s} \right)^{\beta} \left( \frac{\nu_s}{\nu_1} \right)^{\beta-\beta_s}, & \nu_{2}\geq \nu\geq \nu_{H}\\
\left( \frac{\nu}{\nu_s} \right)^{\beta} \left( \frac{\nu_s}{\nu_1} \right)^{\beta-\beta_s}, & \nu_{s}\geq \nu \geq \nu_{2}\\
\left( \frac{\nu}{\nu_1} \right)^{\beta-\beta_s}, & \nu_{1}\geq \nu \geq \nu_{s}\\
1,& \nu \geq \nu_{1}
\end{cases}
\end{eqnarray}
\noindent Here $\nu$ is the today's frequency of the PGWs, $\nu_H=1/\ell_{H}\sim 2\times 10^{-18}$Hz is the Hubble frequency, and the upper limit frequency is chosen $\nu_1=3\times 10^{10}$Hz \cite{zhang2005relic}. By considering $\Omega_{\Lambda}\sim 0.7$ and $\Omega_{m}\sim 0.3$, it turns out that $\nu_{E}\sim 1.5 \times 10^{-18}$Hz and $\nu_2=58.8 \nu_{H}\sim 1.17\times 10^{-16}$Hz \cite{zhang2005relic}. The frequency $\nu_s$, which corresponds to waves that crossed the horizon at the end of the reheating stage, is undetermined and in general depends on the choice of the inflationary parameters $(\beta,\beta_s,\mathcal{A})$. The parameter $\mathcal{A}$ determines the initial spectral amplitude of tensor perturbations and its upper limit is inferred from the constraints on the tensor to scalar ratio, $r\equiv \Delta_{h}^2(K_0)/\Delta_{R}^2 (K_0)$, where $K_0$ is the pivot wave number $0.05$ $\text{Mpc}^{-1}$. Taking $\Delta_{R}^2(K_0)=2.2\times 10^{-9}$ from BOSS DR12 and $r\leq0.044$ from the recent Planck results \cite{ballardini2022new,tristram2021planck}, gives $\mathcal{A}_{-1.9}\leq 4.9\times10^{-4}$ for $\beta=-1.9$ and $\mathcal{A}_{-1.8} \leq 4.2\times 10^{-4}$ for $\beta=-1.8$. Throughout the paper, we mostly consider the upper limits for $\mathcal{A}$ when examining different quantities. Moreover, theoretical considerations put a constraint on the allowed values of $\beta_s$ and $\nu_s$ \cite{zhang2005relic}, 
\begin{eqnarray}\label{eq:6}
1.484\times10^{58}\frac{\mathcal{A}}{(1+z_{E})^3}=\left( \frac{\nu_1}{\nu_H} \right)^{-\beta} \left( \frac{\nu_1}{\nu_s} \right)^{\beta_s},
\end{eqnarray}
\noindent where $z_E$ is the value of redshift at the end of the matter-dominated stage and $1+z_E=\left( \Omega_{\Lambda}/\Omega_{m} \right)^{1/3}\sim 1.33$. Taking $\nu_s=10^{8}$Hz as is most favored in the literature, determines the value of $\beta_s$ \cite{zhang2005relic}. 

Throughout the paper, we use two sets of values for the model parameters $(\beta,\beta_s,\mathcal{A})$ : $(-1.9,0.38,4.9\times10^{-4})$, and  $(-1.8,1.49,4.2\times10^{-4})$. The specific shape of the today's PGWs spectrum together with the corresponding squeezing factor (\ref{eq:5}) in logarithmic scale is shown in Figure. ~\ref{FIG1} for the two sets of parameters.

\subsection{\label{subsec:2.B}Optical medium analogy}

The OMA concept states that the Maxwell's equations in a general curved space-time described by the metric $g_{\mu\nu}$ without external charges and currents, can be replaced by the Maxwell's equations in flat space-time but in the presence of a bi-anisotropic magneto-dielectric media \cite{plebanski1960electromagnetic}. In other words, one can put the effect of geometry (gravity) into a specific material medium. The Cartesian coordinates of the displacement and magnetic fields ($\mathbf{D}$, $\mathbf{H}$) are determined by \cite{plebanski1960electromagnetic,schleich1984general}
\begin{eqnarray}\label{eq:7}
D_{i}(\mathbf{r},t)&=&\varepsilon_{ij}E_{j}(\mathbf{r},t) + (\mathbf{G}\times\mathbf{H}(\mathbf{r},t))_{i},\nonumber\\
B_{i}(\mathbf{r},t)&=&\mu_{ij}H_{j}(\mathbf{r},t) + (\mathbf{G}\times\mathbf{E}(\mathbf{r},t))_{i},
\end{eqnarray}
where the relative permittivity and permeability tensors $(\varepsilon_{ij}, \mu_{ij})$ are determined by
\begin{eqnarray}\label{eq:8}
\varepsilon_{ij}=\mu_{ij}=-\sqrt{g}\frac{g^{ij}}{g_{00}} \quad , \quad G_{i}=-\frac{g_{0i}}{g_{00}}. 
\end{eqnarray}
\noindent Here $i,j$ are spatial indices. Equation (\ref{eq:8}) implies that the corresponding medium is magneto-dielectric, in-homogeneous and anisotropic. We also see from (\ref{eq:7}) that the response is local since no convolution is involved, therefore the medium is non-dispersive and non-absorptive. The vector $\mathbf{G}$ represents the rotation of the reference frame, which vanishes for GWs in TT gauge. The background of weak GWs given by (\ref{eq:1}) induces a correction to the vacuum permittivity tensors according to
\begin{eqnarray}\label{eq:9}
\varepsilon_{ij} = \mu_{ij} &=& \delta_{ij}-h_{ij}(\mathbf{r},t) \nonumber\\
&=& \delta_{ij}- \sum_{\gamma=+,\times} \int\frac{d^3\mathbf{K}}{(2\pi)^{3/2}} e^{\gamma} _{ij}[\hat{\mathbf{K}}] \Big( h_{\gamma}(\mathbf{K})  e^{-i\Omega t } \nonumber\\
&+& h_{\gamma}^{\ast}(\mathbf{K}) e^{i\Omega t } \Big),
\end{eqnarray}
The analogue medium inherits the inhomogeneity and anisotropic features of the GWs background.

\section{\label{sec:3}CLASSICAL MODEL}

\subsection{\label{subsec:3.A}Solution to wave equation}

The Maxwell's equations in the presence of the medium specified by (\ref{eq:9}) can be rewritten as
\begin{eqnarray}\label{eq:10}
&&\nabla\cdot\mathbf{B}=0, \nonumber\\
&&\nabla\cdot\left( \varepsilon(\mathbf{r},t)\mathbf{E} (\mathbf{r},t)\right)=0, \nonumber\\
&&\nabla\times\left( \mu^{-1}(\mathbf{r},t) \mathbf{B}(\mathbf{r},t) \right) =\partial_{t}\left( \varepsilon(\mathbf{r},t) \mathbf{E} (\mathbf{r},t) \right),\nonumber\\
&&\nabla\times \mathbf{E} (\mathbf{r},t) = -\partial_{t}\mathbf{B}.
\end{eqnarray}
In terms of the scalar and vector potentials, one has $\mathbf{E}=-\nabla\phi -\partial_{t} \mathbf{A}$ and $\mathbf{B}= \nabla\times\mathbf{A}$. It can be shown that in order to exclude the extra degree of freedom, i.e., to have $\phi=0$, it is enough to fix the gauge by imposing the condition $\nabla\cdot[ \varepsilon(\mathbf{r},t) \dot{\mathbf{A}} (\mathbf{r},t)]=0$ which gives rise to the following equation of motion for the vector potential 
\begin{eqnarray}\label{eq:11}
\nabla\times\left( \mu^{-1}(\mathbf{r},t)\nabla\times \mathbf{A} \right) = - \varepsilon(\mathbf{r},t) \frac{1}{c^2} \partial^2_{t}\mathbf{A}.
\end{eqnarray}
In this equation, we have neglected a term proportional to $\dot{\varepsilon} \dot{\mathbf{A}}$ in the adiabatic approximation, where $\Omega_{gw}\ll\omega_{em}$. This approximation is practical for the whole frequency range of PGWs. Indeed, this frictional term is responsible for the photon-graviton exchange phenomena, namely, the inverse-Gertsenshtein effect \cite{zel1973electromagnetic}. This effect is particularly important in GW detectors whose physical length is of the order of the GW to be detected, such as the microwave cavities \cite{berlin2022detecting}. However, as far as PGWs are concerned, one may neglect the photon-graviton exchange, due to their far-resonant interaction with optical field.

In order to solve (\ref{eq:11}), one expands the field $\mathbf{A} (\mathbf{r},t)$ over a complete set of mode functions $\mathbf{F}_{k} (\mathbf{r},t)$, \cite{glauber1991quantum,amooshahi2010canonical}
\begin{eqnarray}\label{eq:12}
\mathbf{A}(\mathbf{r},t)=\sum_{k}\sqrt{\frac{\hbar}{2}} \Big( \alpha_{k}\nu_{k}(t)\mathbf{F}_{k}(\mathbf{r},t) + c.c. \Big),
\end{eqnarray}
where $k\equiv(\epsilon,\mathbf{k})$ denotes different modes of the EM field with wave vector $\mathbf{k}$ and polarization $\epsilon$. Here the quantum prefactor $\hbar$ is included intentionally, but at the classical level one can substitute $\alpha_{k}\sqrt{\hbar}\rightarrow \mathcal{A}_{k}$. The time-harmonic behaviour of $\mathbf{A}$ is included in $\nu_{k}(t)$ which recasts to $e^{\pm i\omega_{k}t}$ in the vacuum case. The mode-functions $\mathbf{F}_{k}(\mathbf{r},t)$ contain the spatial factor $e^{i\mathbf{k}\cdot\mathbf{r}}$, and we let it to have a slowly varying envelope as well. The slowly varying envelope approximation (SVEA) implies that \cite{schleich1984general} 
\begin{eqnarray}\label{eq:13}
\Big|\frac{\ddot{\mathbf{F}}_{k}}{\dot{\mathbf{F}}_{k}}\Big|\ll \Big|\frac{\dot{\mathbf{F}}_{k}}{\mathbf{F}_{k}}\Big|\ll \omega_{k},
\end{eqnarray}
By substituting the expansion (\ref{eq:12}) into (\ref{eq:11}) we get
\begin{eqnarray}\label{eq:14}
\nabla\times\left( \mu^{-1}(t) \nabla\times \mathbf{F}_{k} \right) = -\left( \frac{\ddot{\nu}_{k}(t)}{c^2 \nu_{k}(t)}\right) \varepsilon(t) \mathbf{F}_{k},
\end{eqnarray}
By defining $(\ddot{\nu}_{k}/\nu_{k})\equiv -\omega_{k}^{2}(t)$, equation (\ref{eq:13}) becomes an eigenvalue equation for the mode functions $\mathbf{F}_{k}(\mathbf{r},t)$ which satisfy the following orthogonality condition \cite{glauber1991quantum,amooshahi2010canonical}
\begin{eqnarray}\label{eq:15}
\int d^3 r \varepsilon_{ij}(t) F_{ki}^{\ast} F_{k'j} = \delta_{kk'}.
\end{eqnarray}
Due to the local response (\ref{eq:7}) the (temporal) Fourier transformation of the permittivity tensors (\ref{eq:9}) is proportional to delta function, $\delta(\Omega'\pm\Omega)$, which means that the medium is not dispersive and absorptive in the optical range. One can easily check that the following mode functions
\begin{eqnarray}\label{eq:16}
\mathbf{F}_{k}(\mathbf{r},t) = \frac{\hat{\mathbf{u}}e^{i\mathbf{k}\cdot\mathbf{r}}}{\sqrt{m(t)}},
\end{eqnarray}
satisfy the orthogonality condition (\ref{eq:15}) as well as the SVEA condition  (\ref{eq:13}). Moreover, $\hat{\mathbf{u}}$ is the unit polarization vector and $m(t)\equiv \varepsilon_{ij}(t) u_{i}u_{j}$ is a slowly varying function. The gauge condition $\nabla\cdot[ \varepsilon(\mathbf{r},t) \dot{\mathbf{A}} (\mathbf{r},t)]=0$ now implies that $\hat{\mathbf{k}}\cdot (\varepsilon(t)\hat{\mathbf{u}})=0$. Plugging the solution (\ref{eq:16}) in (\ref{eq:14}) leads to
\begin{eqnarray}\label{eq:17}
\mathbf{k}\times\left( \mu^{-1}(t) \mathbf{k}\times \hat{\mathbf{u}} \right) = -\frac{\omega_{k}^{2}(t)}{c^2} \varepsilon(t)\hat{\mathbf{u}}.
\end{eqnarray}
Substituting (\ref{eq:9}) in (\ref{eq:17}) yields to the ordinary and extra-ordinary wave solutions \cite{meschede2017optics}, and the governing equation for the temporal mode functions $\nu_{k}(t)$ outcomes,
\begin{eqnarray}\label{eq:18}
\ddot{\nu}_{k}(t) = -\omega_{k}^{2}(t) \nu_{k}(t),
\end{eqnarray}
where
\begin{eqnarray}\label{eq:19}
\omega_{k}^{2}(t) \equiv \omega_{0}^{2}/n^2(t).
\end{eqnarray}
Here $\omega_{0}=c|\mathbf{k}|$ is the EM frequency at the initial time. The function $n(t)$ defined by (\ref{eq:19}) is the refractive index of the medium which could be found using the Fresnel equation \cite{bini2014refraction},
\begin{eqnarray}\label{eq:20}
n(t) &=& \sqrt{\text{det}(\varepsilon(t))/\varepsilon_{ij} k_{i}k_{j}} = \left( \varepsilon_{ij} k_{i}k_{j} \right)^{-1/2}\nonumber\\
&=& 1+\frac{1}{2}h_{ij}(t)k_{i}k_{j}.
\end{eqnarray}
Here, $k_i$ is the $i$-th component of the EM wave vector. Also, we have used $\text{det}(\varepsilon(t))=1$ at the linear order $\mathcal{O}(h)$. The last equality comes from the definition of the permittivity tensors (\ref{eq:9}). It has to be noted that the time $t$ is the delayed time $t_d$ defined in section~\ref{subsec:2.A}. That the refractive index of the medium is anisotropic, i.e., depends on the EM propagation direction $\hat{\mathbf{k}}$, reflects the fact that the medium is anisotropic and consequently the EM field experiences different refractive indexes depending on its propagation direction.

\subsection{\label{subsec:3.B}Frequency shift}

Now, one can interpret the result (\ref{eq:19}) as the \textit{frequency shift} of the EM field induced by the GWs, as was expected in the TT frame. To show that, we define the frequency shift as the ratio $\mathcal{Z}\equiv (\omega(t)-\omega(t_0)/\omega_0)$, where $\omega(t_0)$ is the EM frequency at an initial time $t_0$, and $\omega_0$ is the EM frequency in the absence of GWs background (note that the definition of $\mathcal{Z}$ coincides with that of \cite{mashhoon1979detection}). With the help of (\ref{eq:9}) and (\ref{eq:20}) one can show that the frequency shift is determined by 
\begin{eqnarray}\label{eq:21}
\mathcal{Z}= \frac{1}{2}\hat{k}_{i}\hat{k}_{j} \left( h_{ij}^{S}- h_{ij}^{R} \right),
\end{eqnarray}
where \textit{S,R} stand for the \textit{sender} and \textit{receiver} points, in the context of time delayed interferometry (TDI). Thus, (\ref{eq:21}) is in full agreement with the frequency shift of the EM field as it propagates through GWs background\cite{estabrook1975response, mashhoon1979detection,mashhoon1982contribution,ciufolini2001gravitational}. As we will  see, in the TT gauge, this frequency shift leads to the modification of the temporal mode-functions $\nu_{k}(t)$.

The time dependent frequency (\ref{eq:19}) should be considered with care. Generally, this points to the possible particle production by the gravitational field, and in the corresponding quantum field theory, this leads to the ambiguity in defining the vacuum state of the EM field \cite{mukhanov2007introduction}. However, one can show that for a slowly varying GWs background, the corresponding Bogoliubov coefficient is zero and the particle production is negligible. Consequently, the WKB prescription can be used to find the adiabatic vacuum \cite{mukhanov2007introduction},
\begin{eqnarray}\label{eq:22}
\nu_{k}(t) = \frac{1}{\sqrt{\omega_{k}(t)}}e^{\pm i \int_{t_{0}}^{t} \omega_{k}(t') dt'}.
\end{eqnarray}
Hence, the final result of the vector potential, $\mathbf{A}(\mathbf{r},t)\propto \nu(t) \mathbf{F}(\mathbf{r},t)$, encapsulates the effect of GWs on the EM field in the TT frame. However, all subsequent results are gauge invariant. For instance, in the proper detector frame (PD), the space-time itself is assumed to be flat, and the effect of GWs is to change the position of the objects, as well as their separations. The movement of these objects leads to the “phase modulation of light”. Thus the overall effect is the phase change, which is a gauge-invariant quantity. We will see later in Section. \ref{sec:4} that the quantum interaction Hamiltonian (\ref{eq:43}) is proportional to the frequency $\omega(t)$, i.e., to the frequency shift $\mathcal{Z}$, which modifies time evolution of the creation/annihilation operators. 
Hence, although the Hamiltonian (\ref{eq:43}) is not a “gauge invariant” quantity, the overall effects concerning the phase of the EM field, such as decoherence mechanism, the quadrature variance, the spectrum and the phase variance, are gauge invariant quantities. 

\subsection{\label{subsec:3.C}Walk-off angle in the small detector limit}

In the following, for the sake of simplicity and without loss of generality, we assume that the wave vector of light lies on the $x-z$ plane and makes angle $\theta$ with the z-axis, thus $\hat{\mathbf{k}}\equiv(\sin\theta,0, \cos\theta)$. In this case, the refractive index (\ref{eq:20}) is given by
\begin{eqnarray}\label{eq:23}
n(t) &=& \Big(1 + h_{11}(t)\cos^2\theta + h_{22}(t) + h_{33}(t)\sin^2\theta \\
&-& h_{13}(t)\sin 2\theta \Big).\nonumber
\end{eqnarray}
Moreover, the identity $\hat{\mathbf{k}}\cdot(\varepsilon(t)\hat{\mathbf{u}})=0$ implies that
\begin{eqnarray}\label{eq:24}
\hspace{-0.5cm}\frac{u_{z}}{u_{x}} =-\tan\theta \Big( 1-h_{11}(t)+h_{33}(t) - h_{13}(t) (\cot\theta -\tan\theta) \Big).
\end{eqnarray}
Here the polarization state of light varies with time (yet $|u_{x}|^2+|u_{z}|^2=1$), as a result of the birefringence medium. 

\begin{figure}[htb]
 \centering
   \includegraphics[%height=50mm,
   width=.9
   \columnwidth]{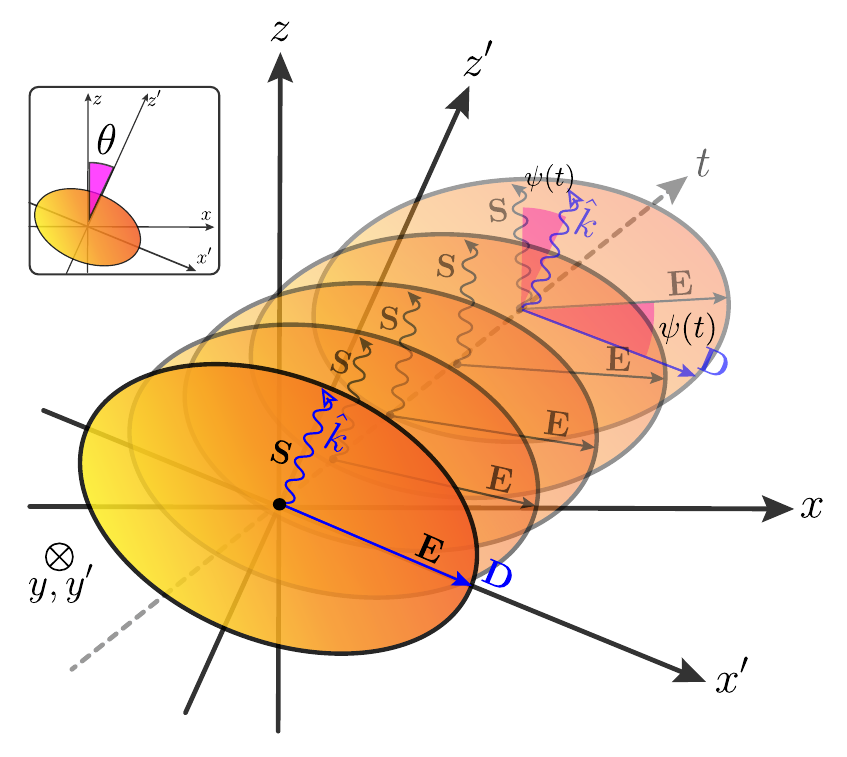}
   \caption{Rotation of the $\mathbf{E}$ and $\mathbf{D}$ at different moments of time. Here ($\hat{x},\hat{y},\hat{z}$) and ($\hat{x}',\hat{y},\hat{z}'$) are the old and new unit vectors, respectively. As light propagates through GWs background, $\mathbf{E}$ and the Poynting vector $\mathbf{S}$ start to rotate, so that the field $\mathbf{E}$ finds a longitudinal component, in the direction of propagation. The angle $\psi(t)$ is the walk-off angle.}
 \label{FIG2}
\end{figure}

The propagation direction of the ray through the medium is determined by the Poynting vector $\mathbf{S}=\mathbf{E}\times\mathbf{H}$. As far as concerns to small detectors, the Maxwell's equations (\ref{eq:10}) imply that (\Rnum{1}) $\mathbf{k}\perp\mathbf{D}$ and (\Rnum{2}) $\mathbf{k}\times \mathbf{H} = -\omega_{0}/n \mathbf{D}$. 
At the initial time $\mathbf{E}$ and $\mathbf{D}$ are parallel. As time elapses, $\mathbf{D}$ stays perpendicular to the propagation direction $\mathbf{k}$ while $\mathbf{E}$ starts to rotate. It turns out that the angle between $\mathbf{E}$ and $\mathbf{D}$, represented by $\psi(t)$, is the same as that of $\mathbf{S}$ and $\mathbf{k}$, which is known as the ``walk-off" angle in birefringence media \cite{meschede2017optics}. Thus the wave vector $\mathbf{k}$ points perpendicular to the wave fronts while the Poynting vector $\mathbf{S}$ shows the direction of energy flux. This situation is illustrated in Figure. ~\ref{FIG2}.
To find the walk-off angle $\psi(t)$, one can rotate the $x-z$ plane by angle $\theta$ and rewrite the polarization vector in the new coordinate system as
\begin{eqnarray} \label{eq:25}
\mathbf{u}'
    %\begin{pmatrix}
    %\cos\theta & 0 & -\sin\theta \nonumber\\
    %0 & 1 & 0 \nonumber\\
	%\sin\theta & 0 & \cos\theta \nonumber\\
	%\end{pmatrix}
	%\begin{pmatrix}
    %u_{x} \nonumber\\
    %u_{y} \nonumber\\
    %u_{z} \nonumber\\
    %\end{pmatrix}\nonumber\\
    =\begin{pmatrix}
    u_{x}\left( \cos\theta-\sin\theta u_{z}/u_{x} \right)\\
    u_{y} \\
    u_{x} \left( \sin\theta+\cos\theta u_{z}/u_{x} \right).
    \end{pmatrix}
\end{eqnarray}
By defining the walk-off angle in the $x'$-direction as $\psi(t) \approx|\tan\psi(t)| = u'_{z'}/u'_{x'}$, and using (\ref{eq:23}) one has
\begin{eqnarray}\label{eq:26}
\psi(t) \approx \left[ h_{11}(t)-h_{33}(t)\right] \sin\theta \cos\theta + h_{13}(t) \cos 2\theta,
\end{eqnarray}
Similarly, we can define the corresponding walk-off angle in the $y$-direction as $\psi_{y}\approx \tan\psi_{y} \equiv u'_{y'}/ ({u'}_{x'}^2+{u'}_{z'}^2)^{1/2}$ which would be zero if one sets $u_{y}=0$ at initial time. Thus GWs act as a birefringence medium that deflect the propagation direction of energy flux. However, as we have seen, the wave vector $\mathbf{k}$ is a constant vector in space time. Equivalently, one could use the perturbed geodesic equation for light rays and show that in the small-detector condition that the gradient of metric perturbations vanishes, there would be no deflection of light and $\mathbf{k}$ is a constant vector.

\subsection{\label{subsec:3.D}Stokes parameters of the EM field in the small detector limit}

As a useful treatment to determine the polarization state of light one may investigate the Stokes parameters, since they are quadratic in the field strength and can be determined through intensity measurements. Using the rotated basis $(\hat{x}',\hat{y},\hat{z}')$ one can define the Stokes parameters as \cite{jackson1999classical}
\begin{eqnarray}\label{eq:27}
s_{0}&=& |\hat{x}'\cdot\mathbf{E}|^2 + |\hat{y}\cdot\mathbf{E}|^2 = a_1^2 + a_2^2 , \nonumber\\
s_{1}&=& |\hat{x}'\cdot\mathbf{E}|^2 - |\hat{y}\cdot\mathbf{E}|^2 = a_1^2 - a_2^2, \nonumber\\
s_{2}&=& 2\textbf{Re}\{ (\hat{x}'\cdot\mathbf{E})^{\ast} (\hat{y}\cdot\mathbf{E}) \} = 2a_1 a_2 \cos(\delta_2-\delta_1), \nonumber\\
s_{3}&=& 2\textbf{Im}\{ (\hat{x}'\cdot\mathbf{E})^{\ast} (\hat{y}\cdot\mathbf{E}) \} = 2a_1 a_2 \sin(\delta_2-\delta_1).
\end{eqnarray}
where we have defined $\hat{x}'\cdot\mathbf{E} = a_1 e^{i\delta_1}$, $\hat{y}\cdot\mathbf{E} = a_2 e^{i\delta_2}$ and $a_1, a_2, \delta_1 , \delta_2$ are real numbers. With the help of (\ref{eq:24}) and (\ref{eq:25}) and assuming $u_{y}=0$ without loss of generality, it turns out
\begin{eqnarray}\label{eq:28}
s_{0}&\approx & s_{1}\approx |u'_{x'}|^2 = u_{x}^2 \Big( 1+ 2[h_{33}(t)-h_{11}(t)]\sin^2\theta \nonumber\\
&+&2h_{13}(t)[\tan\theta-\sin 2\theta] \Big)\nonumber\\
s_{2}&=& s_{3}=0.
\end{eqnarray}
The parameter $s_{0}$ measures the intensity of the wave, the parameter $s_{1}$ gives the preponderance of $x'-$linear polarization over $y-$linear polarization and $s_{2}$ and $s_{3}$ give phase information \cite{jackson1999classical}. Thus (\ref{eq:28}) implies that a $x'-$linear polarized light stays linear (since the $y-$linear contribution is zero). However, as the medium is birefringence, the polarization vector rotates in the $x'-z'$ plane by angle $\psi(t)$, given by (\ref{eq:26}) (see Figure. ~\ref{FIG1}).

\section{\label{sec:4}QUANTUM MECHANICAL MODEL}
\subsection{\label{subsec:4.A}Hamiltonian formalism}

So far we have considered the classical description of light propagation through GWs background. As we aim at detecting quantum nature of GWs, we need to promote our classical treatment to a fully quantum mechanical description where both EM field and GWs (in linearized theory) are quantized properly.

In this section we derive the Hamiltonian of the total system by considering EM and GW fields as two sub-systems coupled to each other using OMA framework. We start by the total Lagrangian density, $\mathcal{L}\equiv \mathcal{L}_{\text{gw}}+\mathcal{L}_{\text{em}}$, where $L = \int d^3r \mathcal{L}$ gives the total Lagrangian of the system. Here,
\begin{eqnarray}\label{eq:29}
\mathcal{L}_{\text{gw}}(\mathbf{r},t) = \frac{c^4}{64\pi G}\left( \frac{1}{c^2} (\dot{h}_{ij})^2 - (\partial_{l}h_{ij})^2 \right),
\end{eqnarray}
describes free dynamics of GWs in the TT gauge \cite{maggiore2008gravitational} and the Lagrangian
\begin{eqnarray}\label{eq:30}
\mathcal{L}_{\text{em}}(\mathbf{r},t ) =\frac{1}{2}\Big( \varepsilon_{ij} (\mathbf{r},t) E^i E^j
- \mu^{-1}_{ij}(\mathbf{r},t) B^{i} B^{j} \Big),
\end{eqnarray}
reproduces the classical Maxwell's equations in the presence of GWs given by (\ref{eq:10}). The permittivity tensors ($\varepsilon,\mu$) specified with (\ref{eq:9}) consist of two terms including the effect of vacuum as well as the GWs. One may rewrite $\mathcal{L}_{\text{em}} \equiv\mathcal{L}_{\text{free}} +\mathcal{L}_{\text{int}}$, where $\mathcal{L}_{\text{free}}$ describes free evolution of the EM field while $\mathcal{L}_{\text{int}}$ describes the EM-GWs coupling.

In the following, we define the corresponding conjugate variable momenta of EM and GW fields, and find the quantized form of the total Hamiltonian. Before doing that, we remind that (\Rnum{1}) $\mathbf{E}=-\nabla\phi -\partial_{t} \mathbf{A}$, $\mathbf{B}=\nabla\times \mathbf{A}$ and the suitable gauge condition $\nabla\cdot[ \varepsilon(\mathbf{r},t) \dot{\mathbf{A}} (\mathbf{r},t)]=0$ excludes $\phi$. Therefore the dynamical degree of freedom of the EM field is the vector potential $\mathbf{A}$, and (\Rnum{2}) as mentioned before, the analogue optical medium is non-absorptive and non-dispersive. Consequently, contrary to the usual procedures of the EM field quantization in the presence of absorptive-dispersive magneto-dielectric media \cite{huttner1992quantization,matloob2004electromagnetic,amooshahi2016canonical}, here we do not treat the medium as a heat bath made of continuum of bosonic modes; Instead, we apply the standard method of quantum field theory to the linearized GWs described by tensor field of rank 2.

\subsubsection{\label{subsec:4.A.1}Quantization of linearized gravitational waves}

Quantization of linearized GWs is usually done within the framework of quantum field theory (QFT), where the GW field is simply treated as a rank-$2$ tensor possessing $2$ different polarization states \cite{dirac1958theory,szczyrba1981hamiltonian, dapor2020modifications,parikh2020quantum}. With the help of the classical Lagrangian (\ref{eq:29}), we define the canonical conjugate momentum of $h_{ij}$, as $\Xi_{ij} \equiv \partial \mathcal{L}_{\text{gw}}/\partial \dot{h}_{ij} =c^2/(32\pi G) \dot{h}_{ij}$. In quantum description, one promotes $h_{ij}$ and $\Xi_{ij}$ to operators,
\begin{eqnarray}\label{eq:31}
\hspace*{-0.5cm}\hat{h}_{ij}(\mathbf{r},t) &=& A \sum_{\gamma=+,\times} \int\frac{d^3\mathbf{K}}{(2\pi)^{3/2}} \frac{e^{\gamma}_{ij}[\hat{\mathbf{K}}]} {\sqrt{2\Omega_{K}}} \Big( \hat{b}_{K} e^{-i(\Omega t -\mathbf{K} \cdot \mathbf{r} )} \nonumber\\
&+& \hat{b}^{\dagger}_{K} e^{i(\Omega t -\mathbf{K} \cdot \mathbf{r})} \Big),
\end{eqnarray}
\begin{eqnarray}\label{eq:32}
\hspace*{-0.5cm}\hat{\Xi}_{ij} (\mathbf{r},t) &=& (-i)\frac{c^2 A}{32\pi G} \sum_{\gamma=+,\times} \int\frac{d^3\mathbf{K}}{(2\pi)^{3/2}}\sqrt{\frac{\Omega_{K}}{2}} e^{\gamma} _{ij}[\hat{\mathbf{K}}] \nonumber\\
& &\Big( \hat{b}_{K} e^{-i (\Omega t -\mathbf{K} \cdot \mathbf{r})}
-\hat{b}^{\dagger}_{K} e^{i(\Omega t -\mathbf{K} \cdot \mathbf{r})} \Big),
\end{eqnarray}
where $K\equiv(\gamma,\mathbf{K})$ and the constant $A$ has to be determined. The bosonic operators $(\hat{b}_{K} ,\hat{b}_{K}^{\dagger})$ are the ladder operators of mode $K$. Following the standard QFT approach, we impose the following equal time commutation relation between $h_{ij}$ and its conjugate momentum,
\begin{eqnarray}\label{eq:33}
\left[ \hat{h}_{ij}(\mathbf{r},t), \hat{\Xi}_{lk} (\mathbf{r}',t)\right] = i\hbar \delta_{il} \delta_{jk} \delta^{(3)}(\mathbf{r}-\mathbf{r}').
\end{eqnarray}
By plugging (\ref{eq:30}, \ref{eq:31}) into (\ref{eq:32}) one gets $A=\sqrt{16 \pi c} \ell_{\text{Pl}}$ \cite{grishchuk2005relic}, where $\ell_{\text{Pl}}=\sqrt{\hbar G/c^3}$ is the Planck length, and the bosonic commutation relation $[\hat{b}_{K},\hat{b}_{K'}^{\dagger}] = \delta_{\gamma\gamma'}\delta^{(3)}(\mathbf{K}-\mathbf{K}')$ outcomes. The free Hamiltonian of GWs turns out to be
\begin{eqnarray}\label{eq:34}
\hat{H}_{\text{gw}} = \frac{1}{2} \int d^3r \left( \frac{32\pi G}{c^2} \Xi_{ij}^2 + \frac{c^4}{32\pi G} (\partial_{l} h_{ij})^2 \right),
\end{eqnarray}
Substituting (\ref{eq:31}, \ref{eq:32}) into the Hamiltonian (\ref{eq:34}) one gets
\begin{eqnarray}\label{eq:35}
\hat{H}_{\text{gw}}= \sum_{\gamma=+,\times}\int d^3K \hbar\Omega_{K}\hat{b}_{K}^{\dagger} \hat{b}_{K}.
\end{eqnarray}

\subsubsection{\label{subsec:4.A.2}Electromagnetic field Quantization}

In section \ref{subsec:3.A} we have specified the mode solutions of the vector potential. We now rewrite the quantized form of (\ref{eq:12}) as
\begin{eqnarray}\label{eq:36}
\hat{\mathbf{A}}(\mathbf{r},t)= \sum_{k} \sqrt{\frac{\hbar}{2}} \left( \hat{a}_{k} \nu_{k}(t) \mathbf{F}_{k}(\mathbf{r},t) + \hat{a}^{\dagger}_{k} \nu^{\ast}_{k}(t) \mathbf{F}^{\ast}_{k}(\mathbf{r},t) \right),
\end{eqnarray}
where the coefficients $\alpha_{k}$ are promoted to operators. The canonical conjugate momentum of $\hat{A}_{i}$ defined by $\hat{\Pi}_{i}(\mathbf{r},t) \equiv \partial\mathcal{L_{\text{em}}}/\partial \dot{\hat{A}}_{i} = \varepsilon_{ij}(\mathbf{r},t) \dot{\hat{A}}_{j} (\mathbf{r},t)$, is given by
\begin{eqnarray}\label{eq:37}
\hat{\Pi}_{i}(\mathbf{r},t)=\varepsilon_{ij} \sum_{k} \sqrt{\frac{\hbar}{2}} \left( \hat{a}_{k} \dot{\nu}_{k}(t) F_{kj} + \hat{a}^{\dagger}_{k} \dot{\nu}^{\ast}_{k}(t) F^{\ast}_{kj} \right).
\end{eqnarray}
Now one can construct the Hamiltonian density by performing the Legendre transform on the Lagrangian density $\mathcal{L}_{em}$. The resulting Hamiltonian is
\begin{eqnarray}\label{eq:38}
\hspace{-0.5cm}\hat{H}_{\text{em}}=\frac{1}{2}\int d^3r \Big( \varepsilon_{ij} ^{-1} (\mathbf{r},t) \hat{\Pi}_{i}\hat{\Pi}_{j} + \mu_{ij}^{-1}(\mathbf{r},t) (\nabla\times \hat{\mathbf{A}})_{i} (\nabla\times \hat{\mathbf{A})}_{j} \Big),
\end{eqnarray}
where we have used the identity $\varepsilon_{ik} \Bar{\Bar {\varepsilon}}_{kl}^{-1}=\delta_{il}$. The standard canonical quantization is done by imposing the following equal time commutation relation between the Cartesian components of the conjugate variables $\hat{\mathbf{A}}$ and $\hat{\mathbf{\Pi}}$
\begin{eqnarray}\label{eq:39}
\left[ \hat{A}_{i}(\mathbf{r},t) , \hat{\Pi}_{j}(\mathbf{r}',t) \right] = i\hbar P_{ij}^{\perp} (\mathbf{r},\mathbf{r}').
\end{eqnarray}
Here $P_{ij}^{\perp}$ is the generalized transverse delta-distribution \cite{glauber1991quantum,amooshahi2010canonical}. By inserting (\ref{eq:36}, \ref{eq:37}) into the canonical commutation relation (\ref{eq:39}) one gets the bosonic commutation relation $[\hat{a}_{k},\hat{a}_{k'}^{\dagger} ] = \delta_{kk'}$ and the other commutators vanish. The resulting Hamiltonian is
\begin{eqnarray}\label{eq:40}
\hat{H}_{\text{em}} &= &\frac{\hbar}{4}\sum_{k} \Big( \hat{a}_{k}\hat{a}_{-k} \left[ \dot{\nu}^{2}_{k} + \omega_{k}^2(t) \nu_{k}^{2} \right] \nonumber\\
&+& \hat{a}^{\dagger}_{k}\hat{a}^{\dagger}_{-k} \left[ \dot{\nu}^{\ast 2}_{k} + \omega_{k}^2(t) \nu_{k}^{\ast 2} \right] \\
& + & \left( 2\hat{a}^{\dagger}_{k}\hat{a}_{k} + \delta^{(3)}(0)\right) \left[ |\dot{\nu}_{k}|^2 + \omega_{k}^{2}(t)|\nu_{k}|^2 \right] \Big).\nonumber
\end{eqnarray}
The Hamiltonian explicitly depends on time and in general, doesn't posses time-independent eigenvectors. Nevertheless, one can still define the instantaneous vacuum at a given moment, $|0_{(t_{0})}\rangle$, to be the lowest energy state of the Hamiltonian $\hat{H}(t_{0})$. This is done by minimizing the expectation value $\langle \hat{H}(t_{0})\rangle$ with respect to the mode functions $\nu_{k}(t_0)$. One can show that the mode functions
\begin{eqnarray}\label{eq:41}
\hspace*{-.6cm}
\nu_{k}(t_{0})=\frac{1}{\sqrt{\omega_{k}(t_{0})}} e^{i\eta_{k}(t_{0})},\dot{\nu}_{k}(t_{0})=i\omega_{k}(t_{0}) \nu_{k}(t_{0}),
\end{eqnarray}
are the preferred mode functions which minimize the Hamiltonian (\ref{eq:40}) and determine the vacuum at a particular moment of time $t_{0}$ \cite{mukhanov2007introduction}. Here $\eta_{k}(t_{0})$ is a function of time which, in the adiabatic limit, coincides with the WKB solution given by (\ref{eq:22}). Thus, for a slowly varying background, $(\Omega_{gw}\ll\omega_{em})$, the WKB approximation implies that the mode functions (\ref{eq:22}) diagonalize the Hamiltonian at all times and the particle creation by the GWs background is negligible, as previously noted by Kip Thorn et al \cite{misner1973gravitation}.

By plugging the mode functions (\ref{eq:22}) in the Hamiltonian (\ref{eq:40}) we obtain the diagonalized Hamiltonian at a given moment of time,
\begin{eqnarray}\label{eq:42}
\hat{H}_{\text{em}} (t)=\sum_{k}\hbar\omega_{k}(t) \hat{a}^{\dagger}_{k} \hat{a}_{k}= \sum_{k}\frac{\hbar\omega_{0k}}{n(t)} \hat{a}^{\dagger}_{k} \hat{a}_{k},
\end{eqnarray}
where $n(t)$ is the refractive index (\ref{eq:20}) and the zero-point energy is excluded. Hence the EM-GWs coupling occurs in the course of frequency shifting (or phase changing) of the EM field. There is two contributions in $\hat{H}_{em}$ describing free evolution of the EM field, as well as its coupling to GWs. The resulting interaction Hamiltonian turns out,
\begin{eqnarray}\label{eq:43}
\hat{H}_{\text{int}} =-\frac{1}{2}\sum_{k}\hbar\omega_{0k}\hat{a}_{k}^{\dagger}\hat{a}_{k} \Big( \hat{h}_{ij}(t) k_{i} k_{j} \Big),
\end{eqnarray}
where $\hat{h}_{ij}(t)$ is given by (\ref{eq:31}). $\hat{H}_{\text{int}}$ represents the interaction between EM and GWs through the intensity dependent coupling, proportional to $\hat{a}^{\dagger} \hat{a}$. This leads to either possible nonlinear interactions $\hat{b}_{K}\hat{a}_{k}^{\dagger}\hat{a}_{k}$ or $\hat{b}_{K}^{\dagger}\hat{a}_{k}^{\dagger}\hat{a}_{k}$, analogous to the three wave mixing process in nonlinear optics and reminiscent of the opto-mechanical coupling in the cavity opto-mechanics \cite{aspelmeyer2014cavity}. In a similar manner, we call this interaction ``opto-gravitational" coupling. As a result, the mutual interaction between the EM field and GWs not only induces EM field dynamics, but also the radiation pressure of the light induces the strain field dynamics, as discussed in a different approach \cite{pang2018quantum}. However, here we consider the continuum of GWs and focus on the quantum GWs' imprints on the EM field observables.

\subsection{\label{subsec:4.B}Quantum dynamics}

Analogue to the cavity opto-mechanical interaction, the intensity-dependent opto-gravitational interaction (\ref{eq:43}) gives rise to EM amplitude-phase noise correlation, and one may expect that this acts as a new source of ponderomotive squeezing (PS) of light, similar to the mechanically-induced PS \cite{fabre1994quantum}. Although the opto-mechanical squeezing of light can not be an evidence of non-classicality of mechanical oscillator, i.e. one can still get squeezing of light without quantizing the mechanical motion, it is shown that ``revivals of optical squeezing" is a pure quantum mechanical effect, in the sense that it exists only if both EM field and mechanical oscillator are quantum entities \cite{ma1811recurrence, ma2020optical}. Following this argument, one may expect that the same happens also for the GWs, and observing the gravitational-induced revivals of optical squeezing could serve as verification of quantum nature of GWs, as was proposed by \cite{guerreiro2020quantum}. In this section we compute the optical variance for a continuum of highly-squeezed PGWs and show that as a result of very small coupling strength between EM-GWs, it takes a very long time to experimentally verify the revivals of optical squeezing. Moreover, the decoherency induced by the continuum of PGWs destroys the EM coherency, which dominates the coherent dynamics. 

The interaction Hamiltonian (\ref{eq:43}) for a single-mode EM field with frequency $\omega_0$ is rewritten as
\begin{eqnarray}\label{eq:44}
\hspace*{-0.5cm}\hat{H}_{\text{int}}/\hbar &=& -\frac{1}{2}\frac{\sqrt{16\pi c^3}}{(2\pi)^{3/2}} \left( \frac{\hbar\omega_0}{E_{\text{Pl}}}\right) \sum_{\gamma=+,\times}\int \frac{d^3K}{\sqrt{2\Omega_{K}}} \nonumber\\
& \times & F_{\gamma}(\hat{\mathbf{K}},\theta) \Big( \hat{b}_{K}e^{-i\Omega t} + \hat{b}^{\dagger}_{K}e^{i\Omega t} \Big) \hat{a}^{\dagger}\hat{a},
\end{eqnarray}
where we have substituted $A=\sqrt{16\pi c} \ell_{\text{Pl}}=\hbar \sqrt{16\pi c^3}/E_{\text{Pl}}$ and the contribution of geometrical configuration of the system is involved in the function $F_{\gamma}(\hat{\mathbf{K}},\theta)$,
\begin{eqnarray}\label{eq:45}
F_{\gamma}(\hat{\mathbf{K}},\theta) \equiv e_{ij}^{\gamma}[\hat{\mathbf{K}}] k_{i}k_{j},
\end{eqnarray}
\noindent Here $\theta$ denotes the spatial angles of the EM field propagation direction (the interferometer arms for instance) with respect to some reference frame. The unitary evolution operator of the total system turns out
\begin{eqnarray}\label{eq:46}
\hspace*{-0.5cm}\hat{U} &=& e^{i E(t)(\hat{a}^{\dagger}\hat{a})^2} e^{\hat{a}^{\dagger}\hat{a} \int d^3K \kappa(\Omega_K) \left(\hat{b}^{\dagger}_{K} \eta_{K}^{\ast}(t)-\hat{b}_{K}\eta_{K}(t)\right)}\nonumber\\
&\times& e^{-i\omega_0 \hat{a}^{\dagger}\hat{a}t} e^{-i\int d^3K \hat{b}_{K}^{\dagger} \hat{b}_{K}(\Omega_{K} t)},
\end{eqnarray}
where $\eta_{K}(t)=1-e^{i\Omega_K t}$, and
\begin{eqnarray}\label{eq:47}
\kappa(\Omega_K)\equiv \frac{1}{4\pi} \left( \frac{\hbar\omega_0}{E_{\text{Pl}}}\right) \left(\frac{c}{\Omega_K}\right)^{3/2} F_{\gamma}(\hat{\mathbf{K}},\theta),
\end{eqnarray}
is a measure of the EM-GW coupling strength. The function $E(t)$ is defined by
\begin{eqnarray}\label{eq:48}
E(t)\equiv\sum_{\gamma}\int d^3 K \kappa^2(\Omega_K) \big(\Omega_K t -\sin\Omega_K t \big).
\end{eqnarray}
Time evolution of the creation and annihilation operators $(\hat{a}^{\dagger},\hat{a})$ is now determined by
\begin{eqnarray}\label{eq:49}
\hspace*{-0.4cm}
\hat{a}(t)=e^{-\int d^3K \kappa(\Omega) \left( \hat{b}_{K}^{\dagger}\eta_{K}(t)-\hat{b}_{K} \eta_{K}^{\ast}(t) \right)} e^{i E(t) (1+2\hat{a}^{\dagger}\hat{a})} \hat{a}
\end{eqnarray}
and it's Hermitian conjugate. The free evolution of the EM field, $e^{-i\omega_0 t}$, is implied in $\hat{a}$. Expression (\ref{eq:49}) is a generalization of the result of \cite{guerreiro2020quantum,bose1997preparation} to the case of a continuum of GWs modes. Squeezed PGWs are denoted by the quantum state $|\{\zeta_{K}\}\rangle$ where the squeezing parameter is $\zeta_{K}=r_{K}e^{i\theta_{\zeta_K}}$ and $(r_{K},\theta_{\zeta_K})$ stand for the squeezing amplitude and phase of mode $K$, respectively. The effect of GWs environment can now be assessed by tracing over the degrees of freedom of the GWs, $|\{\zeta_{K}\}\rangle$. This leads to projection of the EM field operator $\hat{a}$ onto its Hilbert space,
\begin{eqnarray}\label{eq:50}
\hat{a}(t) &\longrightarrow & \langle \hat{a}\rangle = \langle\{\zeta_{K}\}|\hat{a}|\{\zeta_{K}\}\rangle \nonumber\\
&=& \underbrace{e^{-B(t)}}_{\text{decoherent part}} \underbrace{e^{i E(t) (1+2\hat{a}^{\dagger}\hat{a})} \hat{a}}_{\text{coherent part}},
\end{eqnarray}
where the function $B(t)$ is defined by 
\begin{eqnarray}\label{eq:51}
B(t)&\equiv&\sum_{\gamma}\int d^3K \frac{e^{2r_{K}}}{4} \kappa^{2}(\Omega_K) \Big\{ 4\sin^2(\Omega_K t/2)+\cos\theta_{\zeta_K}\nonumber\\
&-&2\cos(\Omega_K t-\theta_{\zeta_K})
+ \cos(2\Omega_K t-\theta_{\zeta_K}) \Big\}.
\end{eqnarray}
From (\ref{eq:50}) one infers that the propagation of quantum light through the GWs background provides two contributions: coherent and incoherent dynamics. The coherent dynamic is driven by the function $E(t)$ which induces coherent phase oscillations. This is the part that contributes to the revivals of optical squeezing \cite{ma1811recurrence,ma2020optical}. The extra-ordinary small coupling strength $\kappa^2 \sim \mathcal{O}(10^{-61})$ present in the expression of $E(t)$ diminish the coherent effect of quantum vacuum, unless one waits for a long time. On the other hand, the incoherent dynamic is driven by the function $B(t)$, and this function has a huge amplification effect stemming from the large squeezing factor $e^{2r_K}\gg1$. Therefore, the highly squeezed nature of PGWs mainly induces incoherent dynamics which has a plausible chance to be detected, as we are going to investigate. We note that in the absence of high squeezing, the factor $e^{2r_K}$ in $B(t)$ is equal to unity, so that both $E(t)$ and $B(t)$ are extremely small. The same argument is true for a vacuum state of GWs, or any quantum state of gravitons possessing inappreciable number of gravitons. Hereafter, we neglect the function $E(t)$ in our consideration and only retain incoherent dynamics induced by $B(t)$.

\section{\label{sec:5}RESULTS}
\subsection{\label{subsec:5.A}Examining $B(t)$ and $e^{-B(t)}$}

It turns out that the decaying factor $e^{-B(t)}$ determines the time evolution of quantum observables of the system. The expression for $B(t)$ is given by (\ref{eq:51}), where the squeezing amplitude of the amplified PGWs is determined by (\ref{eq:5}). The squeezing phase for each mode of the PGWs, $\theta_{\zeta_K}$, is a time-dependent function. This phase is usually considered to be zero in the literature (see e.g., \cite{kanno2019detecsting}). However, as discussed in \cite{grishchuk2001relic}, time evolution of $\theta_{\zeta_K}$ leads to explicit time dependence of the correlation function of the PGWs, $\langle h^2 \rangle$, which implies that we are dealing with a non-stationary process. For ultra-low frequency modes with $\Omega\leq \Omega_{\text{H}}$, the squeezing phase is approximately zero, while for high-frequency modes, $\Omega\gg \Omega_H$, one has $\theta_{\zeta_K}\approx \Omega(t-t_\Omega) \gg 1$, where $t_{\Omega}$ is almost constant during each evolutionary stage. We consider two cases: $(a)$ the usual situation where $\theta_{\zeta_{K}}=0$, and $(b)$ we include the time evolution of the squeezing phase in our calculations. We will see that including $\theta_{\zeta_{K}}\neq0$ generally reduces the decoherence time so that its suppression results in overestimation of the decoherence time (which is the worst case).
\begin{figure}[htb]
\centering
\includegraphics[%height=50mm,
width=0.8 \columnwidth]{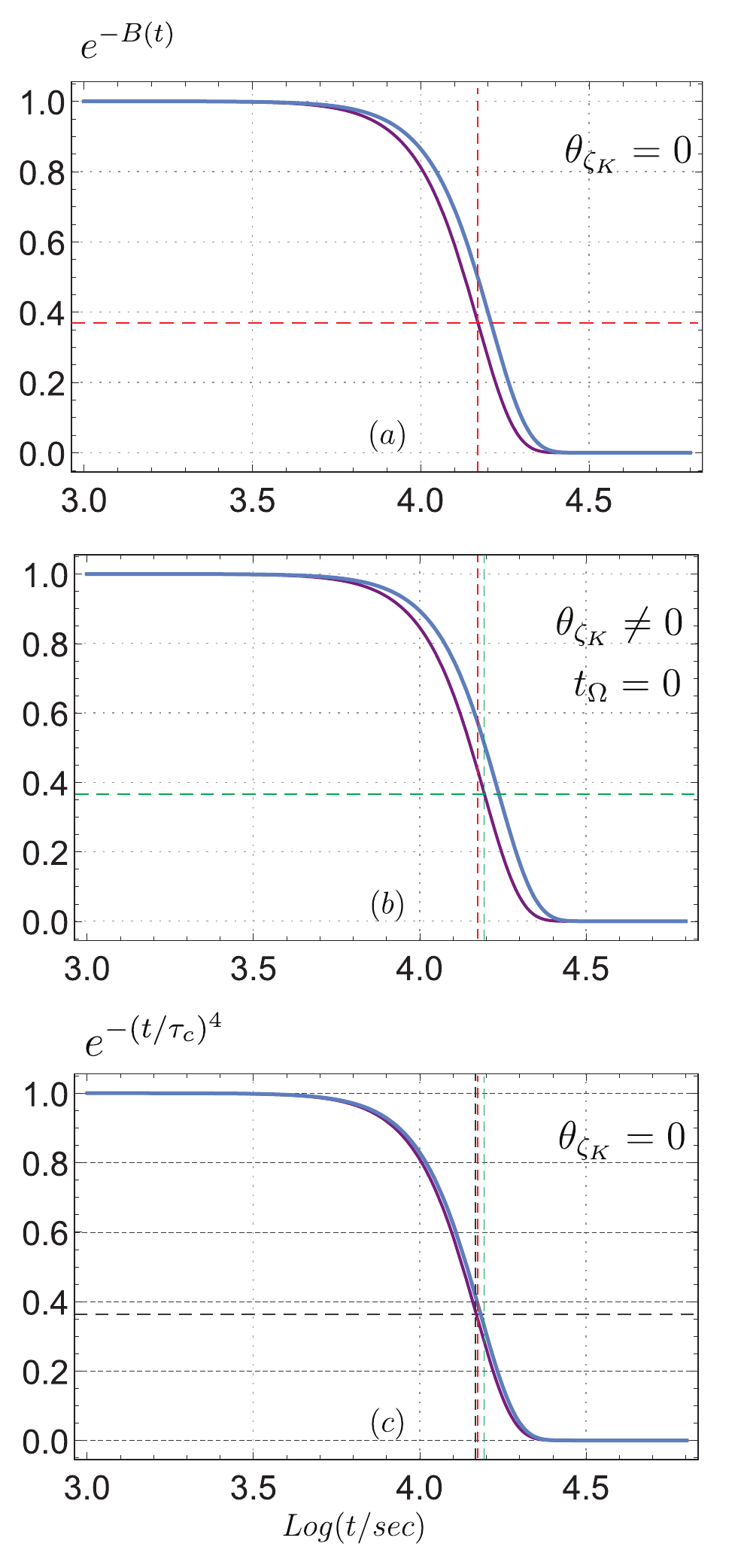}
\caption{ The visibility $e^{-B(t)}$ for various cases: $(a)$ when $\theta_{\zeta_{\Omega}}=0$ based on (\ref{eq:52}), $(b)$ when $\theta_{\zeta_\Omega}\neq0$ and $t_{\Omega}=0$ based on (\ref{eq:54}) and $(c)$ under the linearized approximation where $B(t)=(t/\tau_c)^4$. The purple and blue curves correspond to  $(\beta,\beta_s,\mathcal{A})=(-1.9,0.38,4.9\times10^{-4})$ and $(-1.8,1.49,4.2\times10^{-4})$, respectively. The EM frequency is chosen $\omega_0=1.7\times10^{15}$Hz. It is seen that the linearized approximation works well, hence the analytical expression for decoherence time $\tau_c$ defined by (\ref{eq:55}) gives a plausible estimation of the decoherence mechanism.}
\label{FIG3}
\end{figure}
By setting $\theta_{\zeta_{\Omega}}=0$ in (\ref{eq:51}) and re-arranging terms one obtains
\begin{eqnarray}\label{eq:52}
B(t) = \frac{g(\theta)}{8\pi^2} \left( \frac{\hbar\omega_0}{E_{\text{Pl}}} \right)^2 \int_{\Omega_0}^{\Omega_1} \frac{d\Omega}{\Omega} e^{2r_{\Omega}} \sin^4(\Omega t/2) \Big\vert_{\theta_{\zeta_\Omega}=0},
\end{eqnarray}
where $g(\theta)\equiv \sum_{\gamma} \int d(\Phi_K) d(\cos\Theta_K) F_{\gamma}^2(\hat{\mathbf{K}},\theta)$ accounts for spatial integration. In case of small detectors, it can be disregarded since it's of order unity and is irrelevant. However, for the large detectors where $t\rightarrow t(1-\cos\Psi)$, the spatial factor is of concern. For the PGWs in the frequency range $10^{-18} \leq \Omega\leq 10^{10}$Hz, one can affirm that the major contribution of the integral comes from the ultra-low frequencies which possess the higher squeezing amplitude or graviton intensity (see Figure. ~\ref{FIG1}). Thus, as long as we're concerned with the cumulative effects such as the EM decoherence, we may neglect the contribution of higher frequencies. In this regard, both the ground- and space-based interferometers such as LIGO and LISA, are accounted as small detectors. We proceed by assuming the small detector condition and suppress time delay effects $t(\Psi)\rightarrow t$. 

To include the time-dependency of the squeezing phase, one may consider \cite{grishchuk2001relic}
\begin{eqnarray}\label{eq:53}
\theta_{\zeta_\Omega} = \left\{\begin{array}{lr}
        0 , & \Omega_{0} \leq \Omega \leq \Omega_{\text{H}}\\
        \Omega (t-t_{\Omega}) , & \Omega \geq \Omega_{\text{H}} 
        \end{array}\right\}.
\end{eqnarray}
Inserting (\ref{eq:53}) into (\ref{eq:51}) results in
\begin{eqnarray}\label{eq:54}
B(t) &=& \frac{g(\theta)}{8\pi^2} \left( \frac{\hbar\omega_0}{E_{\text{Pl}}} \right)^2 \bigg\{ \int_{\Omega_0}^{\Omega_{\text{H}}} \frac{d\Omega}{\Omega} e^{2r_{\Omega}} \Big(\sin^4(\Omega t/2) \Big) \\
&+& \int_{\Omega_{\text{H}}}^{\Omega_1} \frac{d\Omega}{\Omega} e^{2r_{\Omega}} \Big( \sin^2(\Omega t/2) \sin^2(\Omega t_{\Omega}/2) \Big) \bigg\} \Big\vert_{\theta_{\zeta_\Omega}\neq 0}\nonumber,
\end{eqnarray}

Generally, the typical order of magnitude of $B(t)$ depends on the value of $t_{\Omega}$, which is a constant for each evolutionary stage of the Universe. Neglecting $t_\Omega$ cancels the contribution of frequencies $\in (\Omega_H,\Omega_1$), and $B(t)$ is underestimated (note that the integrand is non-negative). Thus the function $e^{-B(t)}$ is overestimated, which gives rise to overestimation of the decoherence time $\tau_c$. In this way, we are sure that effects such as the EM decoherence, phase noise, line-width broadening and side-bands are underestimated.

The function $e^{-B(t)}$ for three cases $(a)$ $\theta_{\zeta_{\Omega}}=0$, $(b)$ $(\theta_{\zeta_{\Omega}}\neq 0 , t_{\Omega}=0)$ and $(c)$ under the linearized approximation $\sin\Omega t \sim \Omega t$, is shown in Figure. ~(\ref{FIG3}) for two sets of parameter values. It can be seen that the visibility reaches to $e^{-1}$ at a specific time scale $\sim 10^{4}$. In particular,the decoherency is sensitive to change of the $\mathcal{A}$, i.e., the tensor-to scalar ratio. These plots correspond to $\mathcal{A}\sim 4\times 10^{-4}$, but in general, decreasing $\mathcal{A}$ increases the decoherence time and vice versa.

In order to find the analytical expression for $\tau_c$, one may use the fact that for the low-frequency PGWs, which have the major contribution in the decoherency, $\Omega t\ll 1$ for $t\sim 10^4$sec. In this approximation, one may substitute $\sin\Omega t$ by $\Omega t$ in the expression $B(t)$ given by (\ref{eq:51}). Hence, the function $e^{-B(t)}$ is rewritten as $e^{-(t/\tau_c)^4}$, where the gravitational induced decoherence time $\tau_c$ is defined by
\begin{eqnarray}\label{eq:55}
\tau_c(\beta,\beta_s,\mathcal{A}) \equiv \left( \frac{g(\theta)}{64 \pi^2} \left( \frac{\hbar\omega_0}{E_{\text{Pl}}} \right)^2 \int_{\Omega_0}^{\Omega_1} d\Omega \Omega^3 e^{2r_{\Omega}(\beta,\beta_s,\mathcal{A})}\right) ^{-1/4},
\end{eqnarray}
where the dependency on the inflationary parameters is written explicitly. Now from (\ref{eq:55}) it is clear that the more the squeezing amplitude (gravitons intensity), the faster the decaying behaviour. The linearized approximation is illustrated by panel $(c)$ in Figure. ~\ref{FIG3}, where the decaying time scale matches well with that of exact expressions.
The linear approximation lets us to use the analytic expression $B(t)=(t/\tau_c)^4$ without loss of precision. For the two sets of parameters considered in this study, $\tau_c(-1.9, 0.38, 4.9\times10^{-4})= 1.47\times10^{4}$ sec and $\tau_c( -1.8, 1.49, 4.2\times10^{-4})= 1.51\times10^{4}$ sec. 

\subsection{\label{subsec:5.B}Optical variance}

The quadrature variance of the EM field is defined by $\Delta\hat{\chi}_{\vartheta}(t)\equiv \langle \hat{\chi}^{2}_{\vartheta}(t) \rangle - \langle \hat{\chi}_{\vartheta}(t) \rangle^2$, where the $\vartheta$-dependent quadrature is defined as $\hat{\chi}_{\vartheta}(t) = \hat{a}(t)e^{-i\vartheta} + \hat{a}^{\dagger}(t) e^{i\vartheta}$ and the expectation value involves the trace over the GWs environment, as in (\ref{eq:50}). We take the initial state $|\psi\rangle=|\alpha\rangle\otimes |\{\zeta_K\}\rangle$, where $|\alpha\rangle$ is the coherent state of laser with $\left\vert\alpha\right\vert^2=\bar{n}$ the mean number of photons. The analytical expression for the variance is given by
\begin{widetext}
\begin{eqnarray}\label{eq:56}
\hspace*{-1cm}\Delta\hat{\chi}_{\vartheta}(t)& = &1 + 2 \bar{n} \bigg\{ e^{-\bar{n}\left(1-\cos4E(t)\right)-2B(t)}\cos\left[2\vartheta- 4E(t)-\bar{n}\sin(4E(t)) \right] \nonumber\\
&-& 2 e^{-2\bar{n} \left( 1-\cos2E(t) \right)-B(t)} \cos^2\left[\vartheta- E(t) -\bar{n}\sin(2E(t)) \right] + 1 \bigg\},
\end{eqnarray}
\end{widetext}
Equation (\ref{eq:56}) is a generalization of result of \cite{guerreiro2020quantum} for a continuum of squeezed GWs. In the single mode case considered in \cite{guerreiro2020quantum}, the variance experiences revivals of squeezing, that is to say the variance dives periodically below $1$, which is basically due to the coherent contribution of the vacuum $E(t)$. Nevertheless, as discussed in \cite{ma1811recurrence}, the typical time scale needed to observe the revivals in the opto-mechanical system is $ t_{\text{obs}} \sim T/\kappa^2$, where $T = 2\pi/\Omega$ is the typical mechanical period time and $\kappa$ is the opto-mechanical interaction strength. For the set of parameters related to the quantized GWs, $\kappa^2 \sim 10^{-61}$ that results in an irrationally large observation time to see the revivals. Moreover, the decoherency induced by the continuum of squeezed PGWs ruins the coherent dynamics well before seeing any coherent effect. Hereafter, we neglect the function $E(t)$ (which is proportional to $\kappa^2$) in our consideration and only retain the incoherent dynamics driven by $B(t)$. Equation (\ref{eq:56}) reduces to
\begin{eqnarray}\label{eq:57}
\Delta\hat{\chi}_{\vartheta=0}(t) &=& 1 + 2\bar{n} \left(e^{-B(t)}-1 \right)^2\nonumber\\
&=& 1 + 2\bar{n} \left(e^{-(t/\tau_c)^4}-1 \right)^2,
\end{eqnarray}
which is always greater than 1. In last line we have used the linearized approximation for $B(t)$ (see section~(\ref{subsec:5.A})). This result is regardless of the value of the phase angle $\vartheta$, which only rotates the quadrature components in the phase space. It can be seen from (\ref{eq:57}) that the variance starts from the reference level $\Delta\hat{\chi}=1$ at initial time, and has a sudden increase to and stabilization at $1+2\bar{n}$ after the characteristic time scale $t\sim\tau_{\text{c}}$. As discussed in section~(\ref{subsec:5.A}), $\tau_c$ depends crucially on the inflationary parameters and its maximum (overestimated) value is about $\sim 1.5\times 10^4$ sec for chosen set of parameters. Thus any increase of the variance in this timescale could be witnessed as existence of the highly-squeezed PGWs. We note that in the absence of high squeezing, the factor $e^{2r_K}$ in $B(t)$ is equal to unity, so that both $E(t)$ and $B(t)$ are extremely small and the variance would not be affected by the GWs background. The same argument is true for a vacuum state of GWs, or any quantum state of gravitons possessing inappreciable number of gravitons.

\subsection{\label{subsec:5.C}Optical power spectrum}

\begin{figure}[htb]
\centering
\includegraphics[%height=50mm, 
width=0.9\columnwidth]{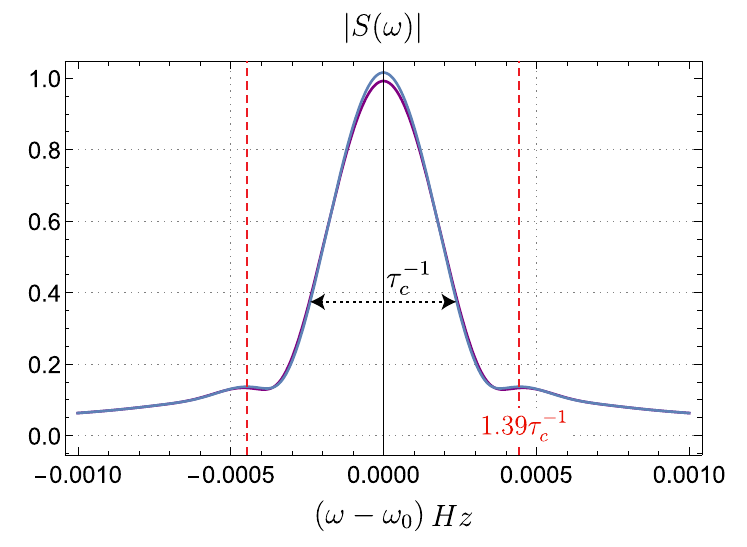}
\caption{The modulus of the optical spectrum, $\vert S(\omega)\vert$, based on (\ref{eq:58}) and (\ref{eq:59}).The blue and purple curves correspond to $(-1.8, 1.49, 4.2\times10^{-4})$ and $(-1.9, 0.38, 4.9\times10^{-4})$, respectively. The laser frequency is taken $\omega_0=1.7\times10^{15}$Hz. The band width broadening is proportional to $\tau_c^{-1}$ Hz, and side bands occur at $\omega \sim \omega_0 \pm 1.39\tau_c^{-1}$Hz.}
\label{FIG4}
\end{figure}

The other quantity of interest is the power spectrum of light, defined by \cite{scully1999quantum}
\begin{eqnarray}\label{eq:58}
S(\omega) = \frac{1}{\pi}\Re \int_{0}^{\infty} d\tau e^{i\omega\tau} G^{(1)}(\mathbf{r},\mathbf{r};\tau),
\end{eqnarray}
where $\Re$ stands for the real part and $G^{(1)}(\mathbf{r},\mathbf{r};\tau)$ is the first degree of coherence of light, say the visibility. For a single mode EM field of linear polarization, using (\ref{eq:49}) and taking the expectation values results in
\begin{eqnarray}\label{eq:59}
\hspace*{-0.5cm} G^{(1)}(\mathbf{r},\mathbf{r};\tau) &=& \frac{\langle \hat{a}^{\dagger}(0)\hat{a}(\tau) \rangle}{\sqrt{\langle \hat{a}^{\dagger}(0) \hat{a}(0)} \sqrt{\hat{a}^{\dagger}(\tau)\hat{a}(\tau) \rangle}} \nonumber\\
&=& e^{-B(\tau)} e^{-\bar{n}(1-e^{2iE(\tau)})} e^{i E(\tau)} e^{-i\omega_0\tau},
\end{eqnarray}
where $B(\tau)$ and $E(\tau)$ are given by (\ref{eq:48}, \ref{eq:51}). Neglecting the small effect of the GWs' vacuum encoded by $E(\tau)$, the resulting effect of PGWs appear as the decaying visibility $e^{-B(\tau)}\sim e^{-(\tau/ \tau_c)^4}$. The power spectrum shows a broadened pick centered at $\omega_0$ and side-bands at $\omega_0 \pm 1.39\tau_c^{-1}$ Hz (Figure. ~\ref{FIG4}). The line-width broadening induced by PGWs is $\gamma=\tau_c^{-1}\sim 6.7\times 10^{-5}$Hz. Not only line-width broadening but also the existence of side-bands in the spectrum are direct consequences of the existence of \textit{squeezed} PGWs. The specific quartic form $e^{-(\tau/\tau_c)^4}$ in the expression of the visibility (\ref{eq:59}) causes the appearance of side-bands and is a unique feature of the squeezed PGWs. The distinction between different gravitons states can be understood from the optical visibility $G^{(1)}(\mathbf{r},\mathbf{r};\tau)$. For to the vacuum, coherent and thermal states of gravitons, one may show that
\begin{eqnarray}\label{eq:60}
G^{(1)}_{\text{vac}} &=& e^{-1/2\int d^3K\kappa^2(\Omega) \sin^2\Omega t},\nonumber\\
G^{(1)}_{\text{coh}} &=& e^{2 i \int d^3K \gamma(\Omega) \sin\Omega t} e^{-1/2\int d^3K\kappa^2(\Omega) \sin^2\Omega t}\\
G^{(1)}_{\text{th}} &=& e^{-\Gamma t}.\nonumber
\end{eqnarray}
\noindent Here $\vert\gamma_{K}\rangle$ represents the coherent state of mode $K$ of GWs, and $\Gamma$ in the last expression is the decoherence rate induced by thermal gravitons, defined in \cite{lagouvardos2021gravitational}. Now, by employing the linear approximation where $\sin\Omega t \sim \Omega t$, it's clear that the corresponding spectrum of (\ref{eq:60}) is a Gaussian or a bell-shape one, without side-bands. Thus, quantum correlations between gravitons in the squeezed state leads to the appearance of side-bands.

To find the position of the side-bands in the spectrum, one may use the asymptotic behaviour of the Fourier transform $\phi(k)=\int_{0}^{\infty}d x e^{ikx}e^{-x^4}$ at $k$ large \cite{boyd2014fourier},
\begin{eqnarray}\label{eq:61}
\hspace*{-0.5cm}\phi(k) & \sim & 2^{7/6}\sqrt{\frac{\pi}{3}}\frac{1}{k^{4/3}} \exp\left({-\frac{3}{16}2^{1/3}k^{4/3}}\right) \\
&\times& \cos\left( \frac{3^{3/2}2^{1/3}}{16}k^{4/3}-\frac{\pi}{6} \right),\nonumber
\end{eqnarray}
\noindent where $k=(\omega-\omega_0)\tau_c$. As it can be seen from Figure. ~\ref{FIG4}, the $k$ large limit is more or less satisfied when $(\omega-\omega_0)\sim 10^{-4}$Hz and $\tau_c \sim 1.5\times 10^4$sec and the location of first side-bands is determined by 
\begin{eqnarray}\label{eq:62}
\omega & = & \omega_0\pm \tau_c^{-1}\left( \frac{\pi/6}{3^{3/2}2^{1/3}/16} \right)^{4/3}\nonumber\\
& \sim & \omega_0\pm 1.39 \tau_c^{-1}.
\end{eqnarray}
%the condition (3^{3/2}2^{1/3} k^{4/3}/16=\pi/6)%
The detuning parameter $\Delta\omega\equiv \omega-\omega_0=1.39\tau_c^{-1}$ which quantifies the separation of side-bands, increases with decreasing $\tau_c$. The magnitude of the decoherence time is governed by the inflationary parameters which are constrained by observational data. A smaller $\tau_c$ needs stronger EM-GW interaction, i.e., higher number of gravitons. The squeezed PGWs possess the highest possible number of gravitons, $\Bar{n}_{\text{gr}}^{\text{pgws}}\sim e^{2r}$, hence the decoherence induced by squeezed PGWs is the strongest possible decoherence mechanism induced by gravitons.
On the other hand, coherent laser beams have finite coherence time which are strikingly smaller than the gravitational induced decoherence time $\tau_c\sim 10^{4}$ sec. This means that the laser field loses its phase correlations well before influenced by the PGWs background. Thus revealing the decoherence induced by gravitons based on the interferometric schema remains a challenge. However, to search for side-bands in the optical spectrum, the fractional-frequency $\Delta\omega/\omega_0\sim(1.39 \tau_c^{-1}/\omega_0)\sim 10^{-19}$ may be touched using ultrastable lasers in near future experiments.

The state-of-the-art fractional-frequency instability of $7\times 10^{-17}$ for $1064$ nm laser along a $2220$ km optical fiber network has been reported recently \cite{schioppo2022comparing}. During passage of light through the network, the phase fluctuations remain extremely small. This opens the room to search for the finest effects in the spectrum of light, including the appearance of side-bands, which is a demonstration of the squeezed PGWs.

The other possibility is to use the ultrastable lasers to search for tiny fluctuations of phase induced by gravitons. 
In the presence of gravitons, the electric phasor executes a random process that finally leads to phase diffusion \cite{lagouvardos2021gravitational}. The phase noise, $\Delta\phi$, can be obtained from the off-diagonal elements of the density matrix, which are basically proportional to the first order correlation function (\ref{eq:59}). It is straight forward to show that the variance is determined by
\begin{eqnarray}\label{eq:63}
\Delta\phi(t)=\sqrt{\langle \phi^2 \rangle} = \left( \frac{t}{\tau_c} \right)^2.
\end{eqnarray}
(One could equivalently follow the Lindblad equation approach to find the density matrix elements, $\rho_{mn}(t)$, or the density distribution function $P(\phi,t)$ \cite{scully1999quantum}). The phase noise increases squarely with time, more faster than the phase change induced by gravitons in thermal state, as considered by \cite{lagouvardos2021gravitational}. The difference comes from the quantum statistical properties of squeezed states, which exclusively bear high correlations of the form $\langle \hat{b}_{K}\hat{b}_{K} \rangle$ and $\langle \hat{b}^{\dagger}_{K}\hat{b}^{\dagger}_{K} \rangle$ that are absent in thermal state. Given $\tau_c\sim 1.5\times 10^{4}$ sec and assuming that coherence time of about $1$ sec is achievable for Hydrogen lasers, the induced phase noise by the squeezed gravitons $\Delta\phi\sim 4.5\times 10^{-9}$ can be measured after a relatively small flight time $\sim 1$sec.

\section{\label{sec:6}Conclusions}

In conclusion, a framework to describe the quantum interaction between GWs background and an EM field is established based on the optical medium analogy. After examining the classical observables of the EM field, the quantized Hamiltonian of the total system is obtained. The imprints of a highly squeezed continuum of GWs on the quantum observables of a laser field, such as the optical variance evolution, the first order coherence function and the spectrum, are studied. The main result is that the GWs background act as a decoherence mechanism that induce an intrinsic line-width broadening on the EM field. The decoherence time crucially depends on inflationary parameters $(\beta,\beta_{\text{s}},\mathcal{A})$. The apparition of side-bands in the optical spectrum is a footprint of the squeezed nature of PGWs. Although experimental search for detecting the optical decoherence induced by the PGWs based on interferometric schema seems challenging, due to smaller intrinsic coherence time of lasers, revealing the side-bands can not only affirm the squeezed nature of PGWs, but also may provide useful information about and put constraint on the inflationary parameters such as ($\beta$, $\beta_{\text{s}},\mathcal{A})$.

\begin{acknowledgments}
The authors wish to acknowledge the conductive support of M. H. Naderi and Arnaud Dupays.
\end{acknowledgments}

%\bibliography{apssamp}  
% ------ bibliography generated by bibtex -----
% ------ bibliography generated by bibtex -----

\providecommand{\noopsort}[1]{}\providecommand{\singleletter}[1]{#1}%
\providecommand{\newblock}{}

\end{document}